\documentclass[sigconf]{acmart}
\AtBeginDocument{%
  }

\usepackage{graphicx}
\usepackage{subcaption} 
\usepackage{booktabs}
\usepackage{diagbox}
\usepackage[most]{tcolorbox}
\usepackage{array}
\usepackage{caption} 
\usepackage{csquotes}
\usepackage{balance}

\copyrightyear{2026}
\acmYear{2026}
\setcopyright{cc}
\setcctype{by}
\acmConference[WWW '26] {Proceedings of the ACM Web Conference 2026}{April 13--17, 2026}{Dubai, United Arab Emirates.}
\acmBooktitle{Proceedings of the ACM Web Conference 2026 (WWW '26), April 13--17, 2026, Dubai, United Arab Emirates}
\acmISBN{979-8-4007-2307-0/2026/04}
\acmDOI{10.1145/3774904.3792799}

\begin{document}

\title{TaoSR-AGRL: Adaptive Guided Reinforcement Learning
Framework for E-commerce Search Relevance}

\author{Jianhui Yang}
\authornote{Equal contribution.}
\email{yangjh23@mails.tsinghua.edu.cn}
\orcid{0009-0004-3547-0472}
\affiliation{%
 \institution{Tsinghua University}
 \city{Beijing}
 \country{China}
 }
\affiliation{%
 \institution{Taobao \& Tmall Group of Alibaba}
 \city{Beijing}
 \country{China}
 }

\author{Yiming Jin}
\authornotemark[1]
\authornote{Corresponding author.}
\email{enxian@alibaba-inc.com}
\orcid{0009-0002-1786-0894}
\affiliation{%
  \institution{Taobao \& Tmall Group of Alibaba}
  \city{Hangzhou}
  \country{China}
  }

\author{Pengkun Jiao}
\email{pkjiao23@m.fudan.edu.cn}
\orcid{0009-0007-0542-3482}
\affiliation{%
  \institution{Fudan University}
  \city{Shanghai}
  \country{China}
  }
\affiliation{%
  \institution{Taobao \& Tmall Group of Alibaba}
  \city{Hangzhou}
  \country{China}
  }

\author{Chenhe Dong}
\email{dongchenhe.dch@alibaba-inc.com}
\orcid{0000-0002-2211-5138}
\affiliation{%
  \institution{Taobao \& Tmall Group of Alibaba}
  \city{Hangzhou}
  \country{China}
  }

\author{Zerui Huang}
\email{huangzerui.hzr@taobao.com}
\orcid{0009-0000-6354-9242}
\affiliation{%
  \institution{Taobao \& Tmall Group of Alibaba}
  \city{Hangzhou}
  \country{China}
  }

\author{Shaowei Yao}
\email{yaoshaowei@taobao.com}
\orcid{0009-0002-3216-7414}
\affiliation{%
  \institution{Taobao \& Tmall Group of Alibaba}
  \city{Hangzhou}
  \country{China}
}

\author{Xiaojiang Zhou}
\email{zxjbupt@163.com}
\orcid{0009-0005-1927-6167}
\affiliation{%
  \institution{Taobao \& Tmall Group of Alibaba}
  \city{Beijing}
  \country{China}
  }

\author{Dan Ou}
\email{oudan.od@taobao.com}
\orcid{0009-0009-9838-5343}
\affiliation{%
  \institution{Taobao \& Tmall Group of Alibaba}
  \city{Hangzhou}
  \country{China}
  }

\author{Haihong Tang}
\email{piaoxue@taobao.com}
\orcid{0000-0002-7103-975X}
\affiliation{%
  \institution{Taobao \& Tmall Group of Alibaba}
  \city{Hangzhou}
  \country{China}
  }

\renewcommand{\shortauthors}{Jianhui Yang et al.}

\begin{abstract}
Query-product relevance prediction is fundamental to e-commerce search and has become even more critical in the era of AI-powered shopping, where semantic understanding and complex reasoning directly shape the user experience and exert an indirect, yet substantial impact on business conversion. Large Language Models (LLMs) enable generative, reasoning-based approaches, typically aligned via supervised fine-tuning (SFT) or preference optimization methods like Direct Preference Optimization (DPO). However, the increasing complexity of business rules and user queries exposes the inability of existing methods to endow models with robust reasoning capacity for long-tail and challenging cases. Efforts to address this via reinforcement learning strategies like Group Relative Policy Optimization (GRPO) often suffer from sparse terminal rewards, offering insufficient guidance for multi-step reasoning, which in turn slows convergence. 

To address these challenges, we propose \textbf{TaoSR-AGRL}, an \textbf{A}dap-tive \textbf{G}uided \textbf{R}einforcement \textbf{L}earning
framework for LLM-based relevance prediction in \textbf{Tao}bao \textbf{S}earch \textbf{R}elevance. TaoSR-AGRL introduces two key innovations: (1) \textbf{Rule-aware Reward Shaping}, which decomposes the final relevance judgment into dense, structured rewards aligned with domain-specific relevance criteria; and (2) \textbf{Adaptive Guided Replay}, which identifies low-accuracy rollouts during training and injects targeted ground-truth guidance to steer the policy away from stagnant, rule-violating reasoning patterns toward compliant trajectories.

TaoSR-AGRL was evaluated on large-scale datasets and through online evaluations on Taobao Search. It consistently outperforms DPO and GRPO baselines, improving relevance accuracy and rule adherence, with measurable gains in user engagement and stable training dynamics. The model has been deployed on Taobao, serving hundreds of millions of users.
\end{abstract}




\maketitle

\begin{figure*}[htbp] 
    \centering
    \includegraphics[width=1.0\textwidth]{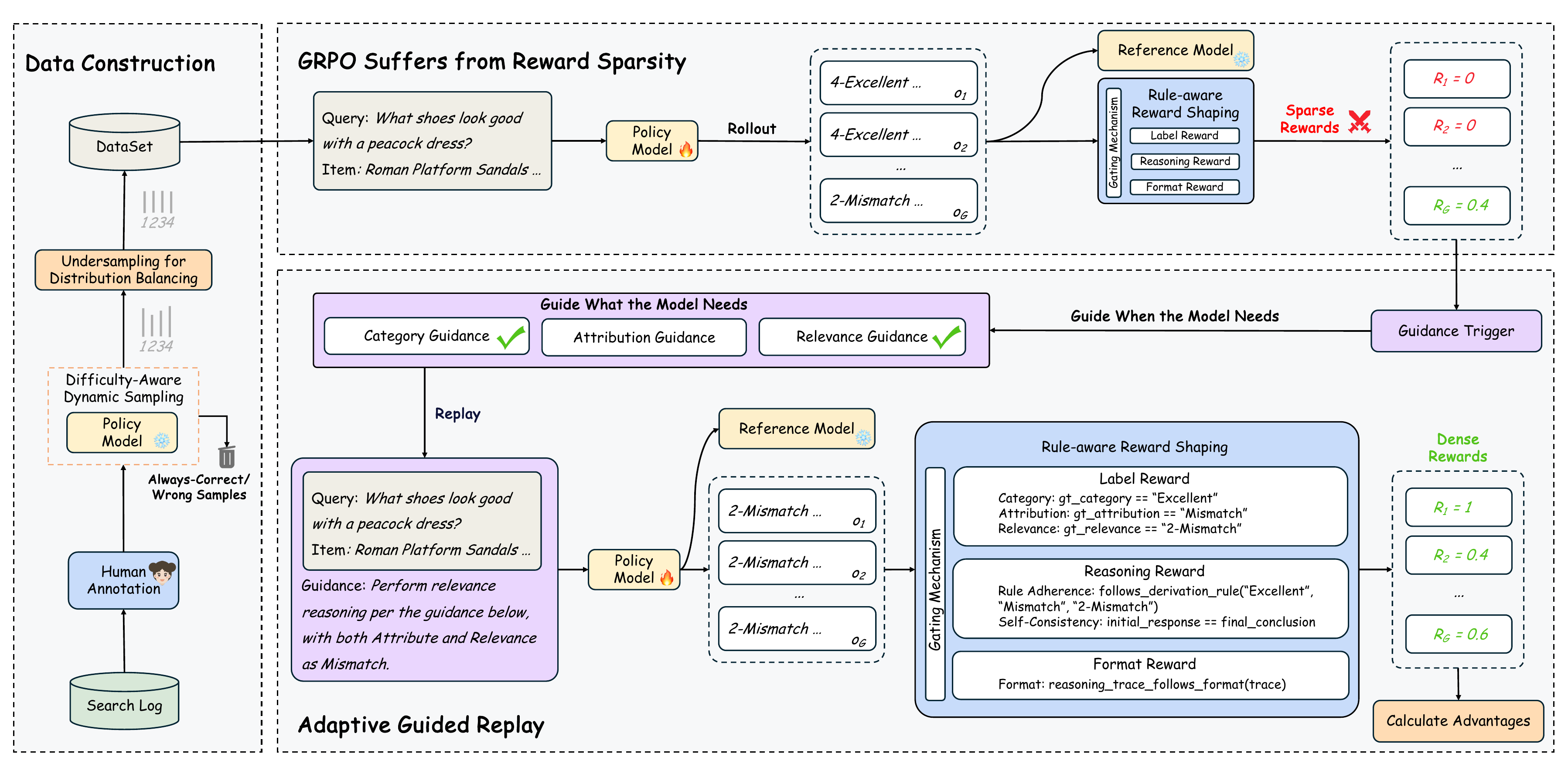}
    \caption{
        Overview of the proposed \textbf{TaoSR-AGRL} for e-commerce search relevance. 
        (1) Data Construction builds a balanced and challenging dataset. 
        (2) Rule-aware Reward Shaping decomposes sparse final rewards into dense, structured signals covering key reasoning steps to ensure faithful and interpretable reasoning. 
        (3) Adaptive Guided Replay selectively replays low-reward, hard samples with targeted guidance to explore higher-value reasoning trajectories.
    }
    \label{fig:overview}
\end{figure*}
\section{Introduction}
In modern e-commerce platforms like Taobao and Amazon, the core task of a search system is to efficiently retrieve and return a highly relevant set of products based on a user's query. Relevance prediction, which quantifies the match between a query and product features, is crucial for both the recall and ranking stages. As e-commerce search shifts towards an AI-powered paradigm focused on semantic understanding and complex reasoning, the importance of relevance prediction is further amplified. Accurate, interpretable, and controllable relevance predictions are becoming the cornerstone of intelligent product retrieval.

Coinciding with this paradigm shift, both e-commerce queries and relevance rule systems are becoming significantly more complex. On one hand, user queries are becoming more open-ended and diverse, such as \enquote{what tea can I drink for a sore throat} or \enquote{snacks suitable for eating in a car}. On the other hand, the platform's relevance rules are increasingly fine-grained and structured. For instance, searching for \enquote{sugared tangerine} and getting \enquote{dried sugared tangerine} is rated as 3-Related, while searching for \enquote{cranberry} and getting \enquote{dried cranberry} is rated as 4-Excellent. These new types of queries demand that models possess rich e-commerce knowledge, the ability to strictly adhere to rules, and multi-step logical reasoning capabilities, posing substantial challenges to traditional search relevance systems.

Large Language Models (LLMs)\cite{yang2025qwen3technicalreport, deepseekai2025deepseekr1incentivizingreasoningcapability, grattafiori2024llama3herdmodels} have demonstrated remarkable reasoning capabilities in complex domains like math\cite{havrilla2024teachinglargelanguagemodels, zhang2024llamaberrypairwiseoptimizationo1like, lai2024stepdpostepwisepreferenceoptimization} and code generation\cite{hui2024qwen25codertechnicalreport, seed2025seedcoderletcodemodel}. This has inspired their application to e-commerce search relevance, where methods like Supervised Fine-tuning (SFT) and Direct Preference Optimization (DPO)\cite{rafailov2024directpreferenceoptimizationlanguage} are used to frame the task as a generative reasoning problem. However, these approaches often exhibit poor generalization on novel, reasoning-intensive long-tail samples, as they primarily learn to mimic patterns within the training distribution rather than acquiring robust, generalizable reasoning ability. To overcome these limitations in generalization and exploration, advanced reinforcement learning (RL) strategies like Group Relative Policy Optimization (GRPO)\cite{shao2024deepseekmathpushinglimitsmathematical} represent a promising direction. Yet, when applied to our context, their effectiveness is severely undermined by the critical challenge of \emph{reward sparsity}\cite{cui2025processreinforcementimplicitrewards, 10.5555/3600270.3600692, ecoffet2019go}. This problem reflects both the intrinsic difficulty of the task and the suboptimal policy initialization: complex long-tail queries demand long, precise chains of reasoning under multiple constraints, making rewarding trajectories inherently rare and difficult to discover; meanwhile, the over-specialization induced by SFT or DPO yields a rigid, narrow exploratory prior, drastically reducing the already low probability of reaching valid reasoning paths. The confluence of these factors leads to exploration stagnation, ultimately crippling both training efficiency and model robustness on challenging unseen instances.

To address the aforementioned issues, we propose an \textbf{A}daptive \textbf{G}uided \textbf{R}einforcement \textbf{L}earning
Framework in \textbf{Tao}bao \textbf{S}earch \textbf{R}elevance (\textbf{TaoSR-AGRL}), a reinforcement learning framework for e-commerce relevance search. Its core idea is to alleviate the exploration bottleneck on hard, long-tail samples and enhance the model's reasoning abilities through rule-aware, fine-grained reward shaping and adaptive guided trajectory resampling. TaoSR-AGRL introduces two key modules. First, Rule-aware Reward Shaping transforms the complex rule-following process into a decomposed, weighted, multi-dimensional reward signal. By explicitly modeling dimensions such as label accuracy, reasoning logic, and formatting constraints, it converts the sparse, binary objective into a dense signal that guides faithful reasoning. Second, Adaptive Guided Replay implements a corrective replay mechanism for samples where the initial rollout fails to produce valid reasoning trajectories. This mechanism identifies the model's weaknesses revealed by Rule-aware Reward Shaping and uses the corresponding ground-truth labels as guidance to perform a second rollout, helping the model explore higher-value reasoning trajectories.

We conduct our research on top of our previous work TaoSR1\cite{dong2025taosr1thinkingmodelecommerce}, using a proprietary relevance-focused large model, TbStar, as the base model. It employs a Mixture-of-Experts (MoE) architecture with 42 billion total parameters and 3.5 billion activated parameters. We construct a test set containing four types of challenging queries to systematically evaluate TaoSR-AGRL. Offline experiments demonstrate that TaoSR-AGRL significantly outperforms baseline methods such as DPO, and GRPO. Furthermore, both online side-by-side human evaluations and A/B testing demonstrate that the framework achieves consistent improvements in retrieval relevance and user engagement in real-world business scenarios. Experiments and ablation studies confirm the effectiveness of Rule-aware Reward Shaping and Adaptive Guided Replay in mitigating reward sparsity on long-tail samples and enhancing the model's reasoning capabilities. The resulting model has been deployed on Taobao, delivering significant business benefits. The main contributions of this paper are as follows:
\begin{itemize}
    \item We propose TaoSR-AGRL, an adaptive guided reinforcement learning framework in Taobao Search Relevance. With Rule-aware Reward Shaping and Adaptive Guided Replay, it effectively alleviates the problems of reward sparsity and exploration stagnation on hard, long-tail samples.
    \item Through comprehensive offline and online experiments, we systematically validate the effectiveness and robustness of TaoSR-AGRL. The resulting model has been successfully deployed on Taobao, further demonstrating the practical value of our method in an industrial setting.
    \item To the best of our knowledge, this is the first work to conduct an in-depth exploration and systematic optimization of reinforcement learning for enhancing LLM reasoning in a large-scale, industrial e-commerce relevance scenarios. It provides a valuable practical example for future research and application in related fields.
\end{itemize}

\section{Related Works}
\subsection{Search Relevance}
The modeling of search relevance has evolved from classical statistical models like TF-IDF \cite{AIZAWA200345} and BM25 \cite{10.1145/1645953.1646237}, which rely on \emph{lexical matching}, to the Learning to Rank (LTR) paradigm. LTR introduced models such as RankNet \cite{10.1145/1102351.1102363} and LambdaMART \cite{Burges2010FromRT}, shifting from handcrafted functions to machine-learned optimization of IR metrics. The deep learning era emphasized \emph{semantic understanding} over lexical overlap. After initial efforts with shallow networks like DSSM \cite{huang2013learning}, the Transformer architecture \cite{vaswani2023attentionneed} and pre-trained models like BERT \cite{devlin-etal-2019-bert} established two dominant architectural patterns: the efficient \emph{bi-encoder} for scalable retrieval \cite{karpukhin-etal-2020-dense} and the high-precision \emph{cross-encoder} for re-ranking \cite{nogueira2020passagererankingbert}, presenting distinct approaches to balance computational efficiency with model effectiveness. More recently, Large Language Models (LLMs) have initiated a paradigm shift, proposing a complementary perspective that frames relevance not merely as a matching problem, but as a reasoning task. Frameworks like LREF \cite{Tang_2025} and TaoSR1 \cite{dong2025taosr1thinkingmodelecommerce} prompt LLMs to reason about e-commerce relevance, leveraging modern alignment techniques such as SFT, DPO \cite{rafailov2024directpreferenceoptimizationlanguage}, and GRPO \cite{shao2024deepseekmathpushinglimitsmathematical}. Despite their power, a critical limitation remains: \emph{reward sparsity} \cite{Zhang2025GRPOLEADAD}. This issue is particularly acute in e-commerce search, as extensive alignment (SFT, DPO) for this singular task can over-specialize the model, constraining the exploratory capacity needed for GRPO to navigate the long-tail distribution. Our work, TaoSR-AGRL, introduces a novel optimization paradigm to specifically address this gap.

\subsection{RLVR for LLM Reasoning}
The paradigm for enhancing complex reasoning in Large Language Models (LLMs) has transitioned from Supervised Fine-Tuning (SFT) to Reinforcement Learning (RL). SFT excels at mimicking demonstrations but suffers from exposure bias and limited generalization \cite{NIPS2015_e995f98d, chu2025sftmemorizesrlgeneralizes}. In contrast, RL fosters superior generalization and exploration through direct environmental feedback \cite{zhang2025onpolicyrlmeetsoffpolicy}. A key development is Reinforcement Learning from Verifiable Rewards (RLVR), which underpins frontier models like OpenAI-o1 \cite{openai2024openaio1card} and DeepSeek-R1 \cite{deepseekai2025deepseekr1incentivizingreasoningcapability}. Algorithms such as GRPO use simple, outcome-based rewards to improve reasoning \cite{zeng2025simplerlzooinvestigatingtamingzero}. However, the efficacy of such methods is substantially limited in \emph{long-horizon tasks} due to \textit{reward sparsity} \cite{nan2025ngrponegativeenhancedgrouprelative, quadros2025llmdrivenintrinsicmotivationsparse, tran2025exploitingtreestructurecredit}. The low probability of discovering a correct solution through random exploration in such problems renders credit assignment highly challenging and hinders learning efficiency.

To address reward sparsity, prior work has explored two main avenues. The first, \textit{process supervision}, densifies rewards by crediting intermediate steps. This is achieved by training Process Reward Models (PRMs) \cite{nye2021workscratchpadsintermediatecomputation, wang2023selfconsistencyimproveschainthought, li2025enhancingreasoningprocesssupervision} or using an \enquote{LLM-as-a-Judge} \cite{Hosseini2024VSTaRTV, Saha2025LearningTP}, though the latter can face robustness and consistency issues \cite{Gu2024ASO}. The second avenue, \textit{external guidance}, simplifies the learning task via hints from a superior LLM \cite{nath2025adaptiveguidanceacceleratesreinforcement, zhang2025critiquegrpoadvancingllmreasoning} or partial reasoning trajectories \cite{zhang2025stephintmultilevelstepwisehints}. Both strategies, however, incur significant computational overhead or rely on external \enquote{oracle} models. In contrast, our work differs by introducing a lightweight, on-demand guidance mechanism. It provides targeted assistance only when needed, eliminating external dependencies and fostering autonomous, efficient learning with minimal computational cost.

\section{Methods}
The e-commerce search relevance task is modeled as a four-class classification problem: 1-Irrelevant, 2-Mismatch, 3-Related and 4-Excellent. Consistent with TaoSR1, the relevance model takes the user query and the item's textual features as input. It outputs a reasoning trajectory in a \enquote{respond-then-think} format, where the relevance label is placed at the beginning of the trajectory to mitigate the effects of error accumulation. Specifically, the model first directly predicts the relevance level for the target task and then generates a complete Chain-of-Thought (CoT) reasoning path to explain the basis for its prediction. The CoT adheres to a predefined reasoning pattern, sequentially deducing the category and attribute levels before determining the final relevance level in accordance with the relevance derivation rules. As e-commerce search transitions towards an AI-powered paradigm, both user queries and rule systems are increasing in complexity. This shift gives rise to challenging long-tail instances that require advanced reasoning to resolve. 

To tackle these cases, we propose \textbf{TaoSR-AGRL}, a reinforcement learning framework for relevance search. The framework comprises two key modules: (1) Rule-aware Reward Shaping, which delivers fine-grained reward signals along the model's reasoning trajectory based on domain-specific relevance criteria, thereby supervising faithful reasoning; and (2) Adaptive Guided Replay, which leverages instance-specific ground-truth labels as precise guidance to resample trajectories for low-accuracy cases, enabling the model to prune the action space and explore higher-value reasoning paths. 
The objective function of TaoSR-AGRL is formulated as follows:
\begin{align}
\mathcal{L}_{\mathrm{TaoSR\!-\!AGRL}}(\theta)
&= \mathbb{E}_{(x,y) \sim \mathcal{D}}
\\ 
&\hspace{-5em}\left[
\frac{1}{G} \sum_{i=1}^G \frac{1}{\left|o_i\right|}
\sum_{t=1}^{\left|o_i\right|}
\min\!\left(
\rho_{i,t}(\theta) \,\hat{A}_{i,t},\ 
\operatorname{clip}\!\left(\rho_{i,t}(\theta),\, 1-\epsilon,\, 1+\epsilon\right) \hat{A}_{i,t}
\right) \right. \nonumber \\
&\left.-\beta D_{\mathrm{KL}}\left(\pi_\theta \| \pi_{\mathrm{ref}}\right)\right] \nonumber \\
\text{s.t.} \quad 
& 0 < \left|\left\{o_i \ \middle|\ \operatorname{is\_equivalent}\!\left(y, o_i\right)\right\}\right| < \gamma
\end{align}

\noindent\textbf{where}:
\begin{equation}
\begin{aligned}
(o_i, R_i) &=
\begin{cases}
\left( \pi_{\theta}(\cdot \mid x),\ R_i^{(0)} \right),
& \text{if } \operatorname{mean}\!\left(\{R_i^{(0)}\}_{i=1}^G\right) > \tau, \\[3pt]
\left( \pi_{\theta}(\cdot \mid x, \mathcal{G}(x,y)),\ R_i^{(\mathrm{rp})} \right),
& \text{otherwise},
\end{cases} \\[6pt]
\rho_{i,t}(\theta) &\triangleq
\frac{\pi_\theta\!\left(o_{i,t} \mid x, o_{i,<t}\right)}
     {\pi_{\theta_{\mathrm{old}}}\!\left(o_{i,t} \mid x, o_{i,<t}\right)}, \quad
\hat{A}_{i,t} \triangleq
\frac{R_i - \operatorname{mean}\!\left(\{R_i\}_{i=1}^G\right)}
     {\operatorname{std}\!\left(\{R_i\}_{i=1}^G\right)}.
\end{aligned}
\end{equation}

Here, $R_i^{(0)}$ and $R_i^{(rp)}$ denote the rewards from the unguided generation and the guided replay, respectively. $\mathcal{G}(x,y)$ represents the targeted guidance, and $\tau$ is the guidance trigger threshold. As illustrated in Figure~\ref{fig:overview}, by adaptively deploying the guidance $\mathcal{G}(x,y)$ for low-reward batches, the model is compelled to learn from expert reasoning. This process enables the discovery of high-value reasoning trajectories, ultimately fostering self-consistent reasoning on challenging long-tail samples.

\subsection{Data Construction}
\label{sec:data_construction}
To train a robust relevance model for complex, real-world scenarios, we constructed a large-scale, high-quality training dataset sourced from user search logs on Taobao. Recognizing that the raw log data suffers from class imbalance, significant noise, and a scarcity of challenging samples, we designed and implemented a sophisticated three-stage data sampling pipeline.

\paragraph{Initial Filtering and Human Annotation}
We first identify four challenging query categories on which existing models perform poorly: negation, affordable alternatives, question-answering (Q\&A), and knowledge-based queries. These initially filtered query logs were then submitted to a professional annotation team for fine-grained annotation based on the query-item relevance. The output of this process is formalized as tuples of \textless \textit{query, item, label\_category, label\_attribute, label\_relevance}\textgreater.

\paragraph{Difficulty-Aware Dynamic Sampling}
To prioritize hard cases that provide the most informative training signal, we adopt a \emph{Difficulty-Aware Dynamic Sampling} strategy, following the general principle in \cite{yu2025dapoopensourcellmreinforcement}. The base model performs multiple rounds of offline prediction over the training dataset, generating diverse predicted trajectories for each instance. Samples that are consistently predicted entirely incorrectly are then subjected to manual inspection, which reveals that their annotation accuracy is considerably low, indicating a substantial level of label noise in this subset. Therefore, our sampling process is designed to remove samples predicted either perfectly correctly or consistently completely incorrectly, as these extreme cases tend to be dominated by label noise or trivial patterns. The retained subset thus consists of medium-difficulty instances, which strike a balance between informativeness and annotation reliability, and are most effective for improving the model's reasoning capacity.

\paragraph{Undersampling for Distribution Balancing}
To mitigate the issue of skewed distribution across different relevance levels, we apply undersampling to the high-frequency classes. This process ultimately creates a dataset that is relatively balanced across all relevance levels, providing a more stable and generalizable sample structure for model training.

\subsection{Rule-aware Reward Shaping}
The fidelity of the intermediate Chain-of-Thought (CoT) is paramount for robust reasoning. However, a critical challenge arises as the aggregation mechanism of derivation rules often masks underlying errors. Consider an instance where the ground truth reasoning involves identifying the category as \texttt{Related} and the attribute as \texttt{Irrelevant}. Under standard protocols, this combination maps to a final \texttt{2-Mismatch} label. A flawed model, however, might correctly classify the category as \texttt{Related} while erroneously labeling the attribute as \texttt{Mismatch}. Crucially, despite this intermediate hallucination, the derivation logic still produces the correct final label. When the model is optimized solely on the reward from this final outcome, it is incentivized to find such shortcuts rather than master the correct reasoning process. This is a classic case of \textit{reward hacking}\cite{skalse2025definingcharacterizingrewardhacking}, yielding two severe consequences. First, it causes impaired generalization because the model fails to learn the genuine reasoning logic, rendering it brittle against out-of-distribution query-item pairs. Second, it leads to reduced trustworthiness and interpretability, as low-quality reasoning chains hinder error attribution and decision traceability in practical applications.

To address these issues, we propose Rule-aware Reward Shaping. Its core idea is to decompose the single, sparse final reward signal into a dense, composite reward that provides fine-grained, multi-dimensional supervision across the entire reasoning chain. More generally, this realizes a transferable principle for rule-driven hierarchical decision tasks, where a final decision is composed from intermediate sub-decisions via explicit rules. In such settings, shaping rewards at each level enables precise credit assignment and discourages shortcut behaviors that exploit rule aggregation. This design ensures that the model not only produces the correct output but also follows a faithful and interpretable reasoning path. Specifically, we formulate the total reward via three orthogonal components:
\begin{itemize}
    \item \textbf{Label Reward}: This component enforces the factual faithfulness of the reasoning chain by anchoring deductions to ground truth. It provides fine-grained supervision for the alignment of predicted category, attribute, and relevance labels, ensuring the final conclusion is factually sound.
    
    \item \textbf{Reasoning Reward}: This component enforces the logical faithfulness of the reasoning process, independent of label correctness. It comprises two sub-mechanisms: (1) \textit{Rule Adherence}, which penalizes step-wise deductions that violate predefined derivation rules; and (2) \textit{Self-Consistency}, which verifies that the final derived conclusion aligns with the initially generated prediction.

    \item \textbf{Format Reward}: This component enforces the structural faithfulness of the output, preserving syntactic integrity. It functions as a binary gating signal, yielding a positive reward solely when the generated CoT strictly adheres to the required schema, thereby filtering out malformed sequences.
\end{itemize}

To prevent the model from earning unearned rewards on flawed or invalid trajectories, we introduce a validity gating mechanism. This gate acts as a prerequisite for all fine-grained rewards, ensuring that credit is only assigned to reasoning paths that meet two fundamental criteria: structural integrity (correct format) and conclusive correctness (correct final relevance). This design establishes a natural \textit{curriculum learning}\cite{soviany2022curriculum}. The model is first compelled to master the primary task of generating well-formed and correct final answers. Only after achieving this baseline competence does it receive the detailed, step-by-step guidance necessary to refine the faithfulness of its internal reasoning process. This staged approach efficiently channels the model's learning, preventing it from getting lost optimizing flawed logic. This principle is formalized in our final reward function $R$:
\begin{equation}
\label{eq:reward} 
R = \mathbb{I}_{\mathrm{Gate}} \cdot 
\left(
w_{\mathrm{cate}} \cdot R_{\mathrm{cate}}
+ w_{\mathrm{attr}} \cdot R_{\mathrm{attr}}
+ w_{\mathrm{reason}} \cdot R_{\mathrm{reason}}
\right)
\end{equation}

\noindent\textbf{where}:
\begin{equation}
\label{eq:gate}
\mathbb{I}_{\mathrm{Gate}} =
\begin{cases}
1, & \text{if } R_{\mathrm{rele}} = 1 \ \land\ R_{\mathrm{format}} = 1, \\
0, & \text{otherwise}.
\end{cases}
\end{equation}

Here, $R_{\mathrm{label}}$ is comprised of three fine-grained rewards: $R_{\mathrm{cate}}$, $R_{\mathrm{attr}}$, and $R_{\mathrm{rele}}$, which assess the correctness of the predicted category, attribute, and final relevance labels, respectively. The gate's state depends on $R_{\mathrm{rele}}$ and $R_{\mathrm{format}}$, which are the binary rewards for the final relevance and format correctness, respectively. The terms $w_{(\cdot)}$ are hyperparameter weights that balance the contributions of the different reward components, providing the flexibility to tailor the training objective by adjusting the emphasis on different facets of the reasoning chain.

\subsection{Adaptive Guided Replay}
While Rule-aware Reward Shaping provides fine-grained feedback, reward sparsity remains a significant challenge, particularly for novel, long-tail queries involving composite rules. 
This issue is best illustrated by the asymmetry in e-commerce search rules. For instance, a \textit{cashmere clothing} item is rated \texttt{4-Excellent} only if its cashmere content exceeds 50\%, whereas for \textit{chiffon}, mere presence suffices. Such composite, asymmetric constraints demand multi-step, fine-grained reasoning, yet models refined via SFT or DPO typically exhibit contracted action spaces and limited exploration. Consequently, when an LLM outputs the correct final label without articulating the intermediate logic, it receives no credit for critical sub-tasks. The resulting learning signal becomes sparse precisely where reasoning is most demanding, thereby degrading credit assignment and hindering convergence on complex e-commerce rules.

To mitigate this issue, we propose an innovative mechanism called Adaptive Guided Replay (AGR). The core idea of this mechanism is to dynamically provide precise guidance during the training rollout phase, based on the model's immediate needs and specific weaknesses. This ensures that the model can learn efficiently from the most difficult and error-prone samples.

\paragraph{Guide When the Model Needs}
To focus limited computational resources on the hard samples that the model most needs to learn from, we first determine when to provide guidance. We hypothesize that samples with low rewards correspond to areas where the model's capabilities are lacking. Specifically, for any sample $x$ in a training batch, we first allow the model to generate an initial result without guidance and calculate its rule-aware reward $R(x)$. When $R(x)$ falls below a guidance trigger threshold $\tau$, we identify it as a sparse-reward sample and initiate the guidance process for it. This way, for \enquote{easy samples} that the model has already mastered, we avoid the overhead of unnecessary guidance and re-computation, thereby significantly improving overall training efficiency.

\paragraph{Guide What the Model Needs}
To prevent the model from engaging in reward hacking or degrading its reasoning abilities when given complete answers, we adopt a \enquote{minimal sufficient guidance} strategy. AGR operationalizes this by performing a fine-grained, on-demand diagnosis. Leveraging the multi-dimensional signals from our rule-aware reward, AGR first calculates the in-batch average reward for each dimension (e.g., category, attribute, and relevance) and compares it against a guidance trigger threshold $\tau$. For any dimension where the reward falls below this threshold, AGR generates targeted guidance $\mathcal{G}(x,y)$ for the selected sparse-reward samples $(x, y)$. This guidance is then formatted into a text prompt and appended to the original input $P(x)$ to form an augmented prompt $P'(x)$. The model performs a \enquote{guided replay} by regenerating its output based on $P'(x)$, allowing it to explore higher-reward reasoning trajectories and fill gaps in its understanding of fine-grained discriminative logic. This dimension-specific intervention effectively mitigates reward sparsity while preserving the model's autonomous reasoning process and suppressing reward hacking.

\section{Experiments}

\begin{table*}[h!]
\centering
\caption{Main results on the \textbf{Balanced Eval Set} and \textbf{In-the-Wild Eval Set}.}
\begin{tabular}{lccccccc}
\toprule
\textbf{Method} & \textbf{Class-1} & \textbf{Class-2} & \textbf{Class-3} & \textbf{Class-4} & \textbf{Good F1} & \textbf{Macro F1} & \textbf{Accuracy}\\
\midrule
\multicolumn{8}{c}{\textbf{Balanced Eval Set}} \\ 
\midrule
TbStar-DPO & 69.30 & 72.86 & 44.50 & 83.04 & 86.02 & 67.43 & 76.94 \\
GRPO & 69.80 & 72.50 & 49.13 & 83.85 & 85.97 & 68.82 & 77.74 \\
GRPO-PR & 69.88 & 73.29 & 47.11 & 83.63 & 86.16 & 68.48 & 77.64 \\
TaoSR-AGRL & \textbf{70.19} & \textbf{73.63} & \textbf{49.28} & \textbf{83.94} & \textbf{86.18} & \textbf{69.26} & \textbf{78.03} \\
\midrule
\multicolumn{8}{c}{\textbf{In-the-Wild Eval Set}} \\ 
\midrule
TbStar-DPO & 44.49 & 63.26 & 45.95 & 82.59 & 86.38 & 59.07 & 73.70 \\
GRPO & 45.41 & 62.95 & 45.93 & 82.72 & 86.49 & 59.25 & 73.83 \\
GRPO-PR & 45.49 & 63.56 & 45.08 & 82.76 & 86.59 & 59.22 & 73.95 \\
TaoSR-AGRL & \textbf{52.77} & \textbf{64.34} & \textbf{46.22} & \textbf{82.79} & \textbf{86.60} & \textbf{61.53} & \textbf{74.28} \\
\bottomrule
\end{tabular}
\label{tab:offline_results}
\end{table*}

\subsection{Experimental Setup}

\paragraph{Dataset and Evaluation Metrics}
The dataset was sampled from the online search logs of Taobao. To comprehensively evaluate the model's performance in complex semantic scenarios, we selected four challenging query categories: negation, affordable alternatives, question-answering (Q\&A), and knowledge-based queries. The training data was obtained through the three-stage filtering process described in Section~\ref{sec:data_construction}. In the initial filtering stage, millions of samples were selected from the raw logs. Subsequently, using Difficulty-Aware Dynamic Sampling, we removed samples that were either all correct or all incorrect, resulting in hundreds of thousands of samples. From this, we further applied Undersampling for Distribution Balancing to adjust the proportions of different relevance levels, ultimately yielding a training set of tens of thousands of samples.

To more comprehensively assess model performance, we constructed two test sets with different distributions. The Balanced Eval Set (B-Eval) maintains a relatively balanced ratio across four query categories. It is designed to validate the model's performance on a balanced sample distribution. The In-the-Wild Eval Set (W-Eval) has a label distribution that approximates the real online environment and is used to examine the model's robustness and generalization capabilities in practical business scenarios.

For offline evaluation, we report per-class F1 for Class-1 (1-Irrelevant), Class-2 (2-Mismatch), Class-3 (3-Related), and Class-4 (4-Excellent), along with Good F1, Macro F1, and Accuracy. Specifically, Good F1 is computed by collapsing Class-3 and Class-4 into a single \texttt{Good} category. This aligns our evaluation protocol with the online business objective, where these high-relevance levels are treated as a unified \texttt{Good Tier}. To validate the real-world impact of our model beyond offline metrics, we conducted online side-by-side human evaluations. This rigorous process is designed to capture user-facing performance differences by directly assessing model outputs on three principal metrics:
\begin{itemize}
    \item \textbf{Good/Same/Bad (GSB):} This metric quantifies the relative performance gain of our model via a pairwise comparison. For the same query, human annotators are presented with two Search Engine Result Pages (SERPs) side-by-side—one from our model and one from the baseline—and label our model's result as \texttt{Good} (superior), \texttt{Same} (equivalent), or \texttt{Bad} (inferior). A GSB score of +$x$\% signifies that our model produced a superior SERP for $x$\% more queries compared to the baseline.
    \item \textbf{Query Goodrate:} This is a holistic, SERP-level metric assessing the absolute quality of search results. Annotators classify the overall utility of an entire SERP into \texttt{Good}, \texttt{Mid}, or \texttt{Bad} tiers. The final metric is the percentage of queries whose SERPs are judged as either \texttt{Good} or \texttt{Mid}, providing a high-level measure of user satisfaction per query.
    \item \textbf{Item Goodrate:} This item-level metric measures the density of highly relevant content on a SERP. It is defined as the proportion of items rated as highly relevant (i.e., \texttt{4-Excellent} or \texttt{3-Related}). The final score is computed by averaging this proportion across all evaluated queries, offering a granular assessment of absolute performance at the item level.
\end{itemize}

\paragraph{Implementation Details}
All experiments related to TaoSR-AGRL were implemented based on the open-source reinforcement learning framework ROLL\cite{wang2025reinforcementlearningoptimizationlargescale}. We used a training batch size of 64, a maximum input length of 2048, and a maximum generation length of 4096. The learning rate was set to 1e-6, with 16 gradient accumulation steps. For each input sample, 16 candidate responses were generated using a sampling temperature of 0.99 and a top-k of 100. For difficulty filtering, an online difficulty threshold range of [0.01, 0.9] was used. To enhance training stability, the PPO clipping range was set to $1 \pm 0.2$, value function clipping to 0.5, and Advantage value clipping to $\pm 2.0$. The guidance trigger threshold $\tau$ was 0.1, and the hyperparameter weights $w_{\text{cate}}$, $w_{\text{attr}}$, and $w_{\text{reasoning}}$ were set to 0.4, 0.4, and 0.2 through preliminary experiments to balance different reward components effectively.

\subsection{Baselines}
For a fair comparison, all methods are benchmarked on the TbStar-DPO backbone. We chose this DPO-aligned model over its conventional SFT counterpart, as our preliminary experiments indicated it establishes a higher performance ceiling, thus providing a more robust evaluation baseline. TbStar is Taobao's proprietary e-commerce MoE model (42B total/3.5B active parameters) with strong domain coverage. We evaluate TaoSR-AGRL against the following representative baselines:
\begin{itemize}
    \item \textbf{GRPO:} A leading critic-free reinforcement learning algorithm that computes advantages using intra-group relative rewards.We used it for a third stage of training on top of TbStar-DPO to enhance the model's reasoning capabilities.
    \item \textbf{GRPO-PR:} This baseline adapts GRPO by augmenting its outcome-based reward with process-level supervision. The total reward is thus formulated as a carefully calibrated weighted sum of relevance, category, and attribute correctness. The coefficients were optimized through an extensive hyperparameter search, resulting in the following reward function:
    \begin{equation}
       R_{GRPO_{PR}} = 0.4 \cdot R_{\text{rele}} + 0.3 \cdot R_{\text{cate}} + 0.3 \cdot R_{\text{attr}}
    \end{equation}
    \item \textbf{TaoSR-AGRL:} Our full proposed method, which integrates Rule-aware Reward Shaping with the Adaptive Guided Replay mechanism to optimize reasoning on complex queries.
\end{itemize}

\subsection{Offline Evaluation}
The main offline evaluation results are presented in Table~\ref{tab:offline_results}. A key finding is that while the standard GRPO algorithm provides some improvement over the DPO-finetuned TbStar model, it quickly hits a performance bottleneck in complex e-commerce search scenarios. On the W-Eval dataset, GRPO and its variant GRPO-PR only achieve a marginal Macro-F1 improvement of 0.2 pt over TbStar-DPO.

In contrast, TaoSR-AGRL successfully breaks through this performance ceiling, achieving new state-of-the-art results on both test sets. On the B-Eval, TaoSR-AGRL achieves a Macro-F1 score of 69.26. Although TaoSR-AGRL was trained on a class-balanced dataset, it demonstrates a more significant advantage on the W-Eval, which mirrors the real-world online distribution. Its Macro-F1 score reaches 61.53, widening the gap with the next-best baseline to 2.28 pt and showcasing strong robustness. Notably, Class-1 improves by 7.28 pt, indicating more calibrated rejection, reduced over-matching, and better class-boundary separation under distribution shift—thereby validating the effectiveness of the framework.

\subsection{Ablation Study}
\begin{figure*}[htbp]
    \centering
    \begin{subfigure}[b]{0.32\textwidth}
        \centering
        \includegraphics[width=\linewidth]{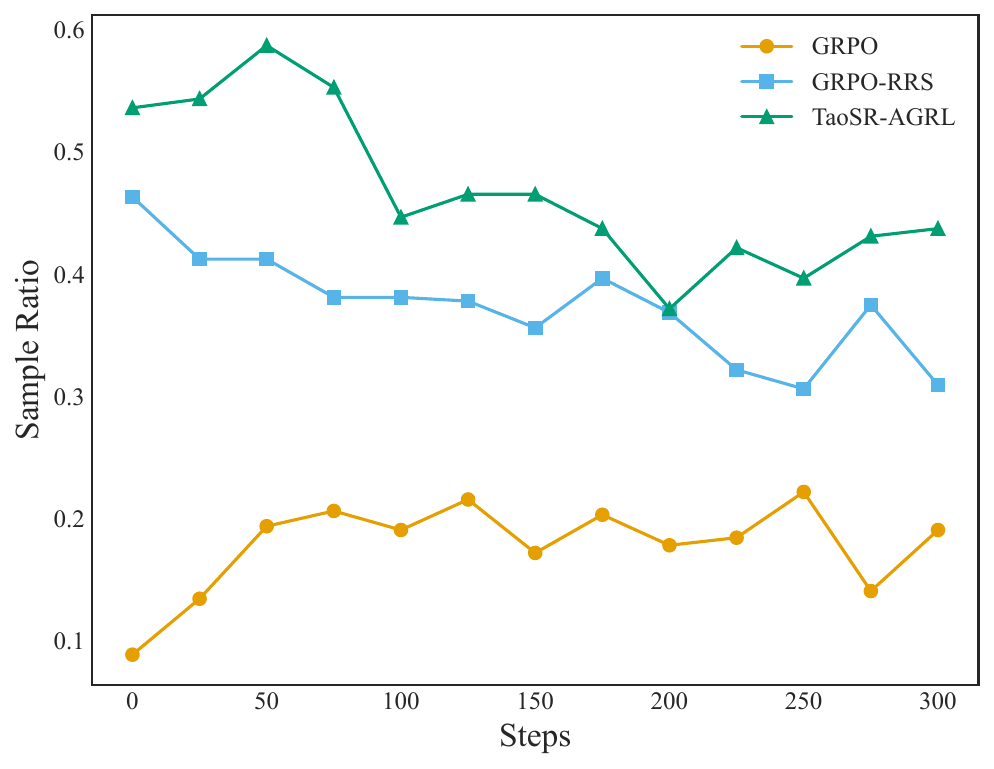}
        \caption{Gradient-Contributing Sample Ratio.}
        \label{subfig:sample_ratio}
    \end{subfigure}
    \hfill
    \begin{subfigure}[b]{0.32\textwidth}
        \centering
        \includegraphics[width=\linewidth]{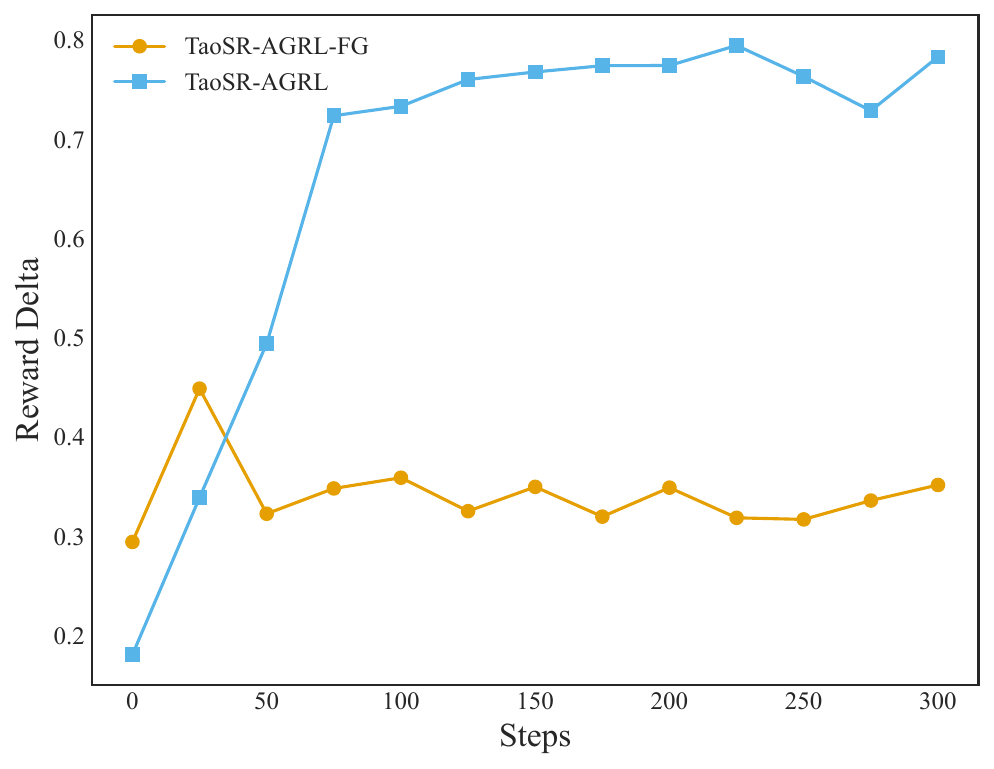}
        \caption{Reward Delta Evolution.}
        \label{subfig:reward_delta}
    \end{subfigure}
    \hfill
    \begin{subfigure}[b]{0.32\textwidth}
        \centering
        \includegraphics[width=\linewidth]{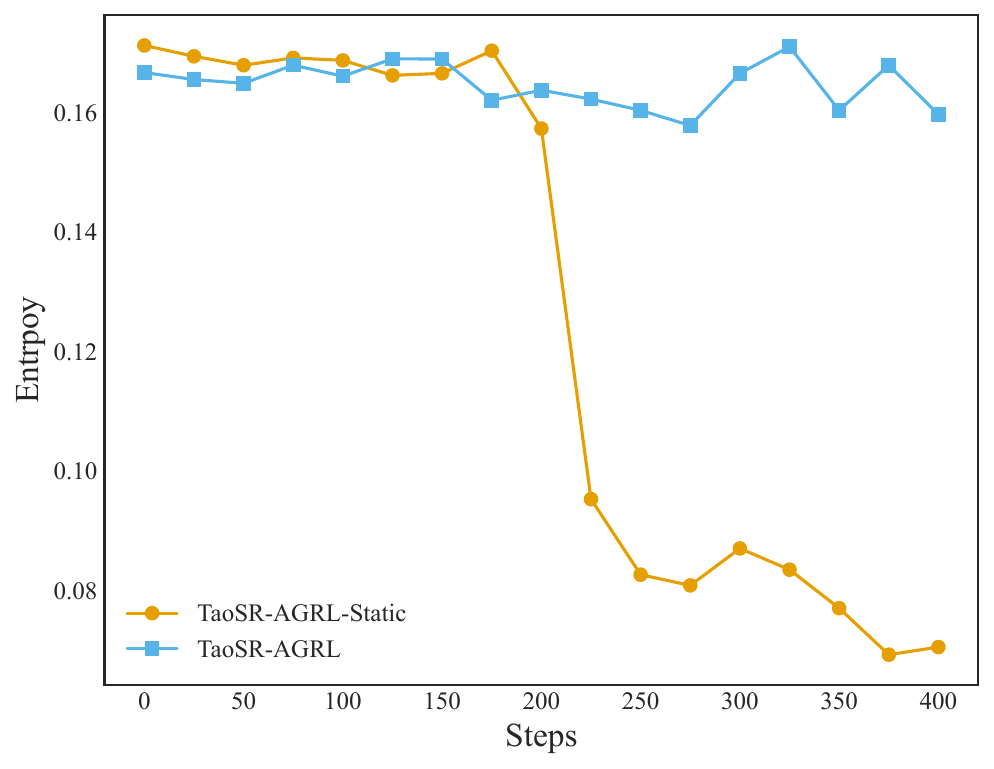}
        \caption{Entropy Evolution.}
        \label{subfig:entropy}
    \end{subfigure}

    \caption{
    Ablation on training dynamics: Compared to its variants, TaoSR-AGRL exhibits (a) higher sample efficiency, (b) greater reward delta, and (c) stable policy entropy preservation, effectively preventing collapse.
}
    \label{fig:ablation}
\end{figure*}

\begin{table*}[htbp]
\centering
\caption{Ablation study results on the In-the-Wild evaluation set.}
\label{tab:ablation_results}
\begin{tabular}{lccccccc}
\toprule
\textbf{Method} & \textbf{Class-1} & \textbf{Class-2} & \textbf{Class-3} & \textbf{Class-4} & \textbf{Good F1} & \textbf{Macro F1} & \textbf{Accuracy}\\
\midrule
GRPO & 45.41 & 62.95 & 45.93 & 82.72 & 86.49 & 59.25 & 73.83 \\
GRPO-RRS & 46.01 & 63.79 & 45.29 & 82.79 & 86.55 & 59.47 & 74.08 \\
GRPO-FG & 48.34 & 64.02 & 43.32 & 82.72 & 86.42 & 59.60 & 74.01 \\
TaoSR-AGRL-FG & 51.28 & 62.77 & 45.79 & 81.22 & 85.42 & 60.00 & 72.41 \\
TaoSR-AGRL & \textbf{52.77} & \textbf{64.34} & \textbf{46.22} & \textbf{82.79} & \textbf{86.60} & \textbf{61.53} & \textbf{74.28} \\
\bottomrule
\end{tabular}
\end{table*}

We conducted a series of ablation studies on the W-Eval benchmark to systematically analyze the performance contributions of each component in TaoSR-AGRL. The main results are summarized in Table~\ref{tab:ablation_results}. In our analysis, the suffix \texttt{-RRS} denotes the variant equipped with \textbf{R}ule-aware \textbf{R}eward \textbf{S}haping. The suffix \texttt{-FG} (\textbf{F}ixed \textbf{G}uidance) denotes a special case of our Adaptive Guided Replay where the guidance is fixed to the ground-truth relevance labels.

\paragraph{Synergy of Gating and Reasoning Rewards.}
To validate the design of Rule-aware Reward Shaping, we ablated its two pillars: the validity gate and the reasoning reward ($R_{\mathrm{reason}}$). Beyond end-task metrics, we use the Rule Adherence Rate (RAR) to measure if the model's reasoning is faithful to its own intermediate steps. Ablating the hard validity gate—either by removing or softening it—introduces reward leakage into invalid trajectories. This not only degrades end-task performance but also lowers RAR, thereby confirming that a strict gate is essential for providing a clean, focused training signal. More critically, removing $R_{\mathrm{reason}}$ causes a steep decline in RAR while leaving end-task metrics largely intact. This result provides clear evidence of reward hacking: the model learns to find the correct answer via logically flawed shortcuts. This demonstrates the synergy between our components: the hard gate ensures learning is confined to successful outcomes, while $R_{\mathrm{reason}}$ ensures the reasoning process leading to those outcomes is logically sound and faithful. A detailed breakdown of these ablations is available in Appendix~\ref{sec:ablation_reward_shaping}.

\paragraph{Impact of Adaptive Guidance on Sample Efficiency.}
The effectiveness of GRPO relies on sampling correct reasoning paths. However, for a monolithic task like relevance search, where the entire reasoning process yields a single outcome, models fine-tuned with SFT and DPO often exhibit a limited exploration space. This policy confinement severely restricts the gradient-contributing sample ratio, as many trajectories contain spurious reasoning steps despite reaching the correct final answer. GRPO-RRS validates this by significantly increasing the ratio of gradient-contributing samples, exposing a high prevalence of such \enquote{false positives} in the baseline. Building on this, TaoSR-AGRL further enhances this ratio, as shown in Figure~\ref{subfig:sample_ratio}. We attribute this to the adaptive guidance replay mechanism, which specifically identifies and provides targeted assistance on challenging instances that the model would otherwise fail on, thereby steering exploration towards more productive reasoning pathways and boosting sample efficiency.

\paragraph{Effectiveness of Adaptive Guidance.}
To isolate the benefit of the adaptive guidance mechanism, we implement TaoSR-AGRL-FG, a variant with ground relevance label as guidance. As reported in Figure~\ref{subfig:reward_delta}, TaoSR-AGRL achieves a significantly higher reward delta within the same training duration. This finding highlights that our adaptive guidance, tailored to the model's evolving weaknesses, is markedly more effective at targeting and rectifying \enquote{reasoning blind spots}. Consequently, this targeted adaptation accelerates convergence and enhances performance in complex inference scenarios.

\paragraph{Preservation of Policy Entropy.}
In reinforcement learning, policy entropy is a critical measure of exploration. High entropy is essential for deep reasoning tasks, as it encourages the model to explore a diverse set of inferential strategies. We compare TaoSR-AGRL against TaoSR-AGRL-Static, a variant that applies guidance indiscriminately to all samples. As illustrated in Figure~\ref{subfig:entropy}, TaoSR-AGRL consistently maintains a higher policy entropy. This suggests that excessive or non-adaptive guidance can lead to \emph{policy collapse}\cite{juliani2024studyplasticitylossonpolicy}: the model becomes overly reliant on the guidance, causing its output distribution to become over-confident and collapse into a narrow set of familiar responses. This dependency ultimately stifles the model's exploratory capabilities and impairs its ability to generalize.

\paragraph{Impact of the Replay Trigger Threshold $\tau$.}
To determine the optimal frequency of intervention, we analyze the model's overall performance across different values of the replay trigger threshold $\tau$. As detailed in Appendix~\ref{sec:appendix_tau}, performance peaks at $\tau=0.1$. This result highlights a fundamental trade-off between corrective intervention and policy autonomy. A low $\tau$ provides insufficient correction for challenging instances, hindering learning. Conversely, a high $\tau$ leads to over-reliance on external guidance, risking the \textit{policy collapse}  discussed previously. This over-reliance can impair the model's intrinsic reasoning ability, even on simpler tasks.

\subsection{Online Evaluation}

\begin{table}[htbp]
\centering
\caption{Online side-by-side human evaluations.}
\begin{tabular}{lccc}
\toprule
\textbf{Category} & \textbf{GSB} & \textbf{Query Goodrate} & \textbf{Item Goodrate} \\
\midrule
Q\&A & +10.57\% & +3.49 pt  & +4.85 pt  \\
Alternative & +5.08\% & +0.76 pt  & +2.85 pt \\
Negative & +2.14\% & +0.55 pt  & +3.64 pt  \\
Knowledge & +10.66\% & +2.14 pt & +3.63 pt \\
\bottomrule
\end{tabular}
\label{tab:online_results}
\end{table}

We validated real-world effectiveness via large-scale online side-by-side human evaluations in production. TaoSR1 served as the baseline bucket and TaoSR-AGRL as the test bucket. We sampled 2{,}000 authentic user queries and compared the top-10 retrieved results from both buckets in an item-level head-to-head comparison. As shown in Table~\ref{tab:online_results}, TaoSR-AGRL consistently outperformed the baseline bucket, achieving average gains of +7.11\% in GSB, +1.74 pt in Query Goodrate, and +3.74 pt in Item Goodrate. Improvements are most pronounced for reasoning-intensive traffic (e.g., Q\&A and knowledge-based queries), indicating that the proposed model effectively integrates e-commerce domain knowledge with rule-driven reasoning to better resolve complex user intent and specific constraints.

To ensure production readiness for high-throughput serving, we implemented engineering optimizations targeting stable response under load, improved resource utilization, and reduced end-to-end latency. Specifically, we optimized model-level load balancing to mitigate traffic skew, reduced cross-datacenter scheduling overhead to lower tail latency, and applied model quantization to accelerate inference while maintaining ranking quality. With these optimizations, Model FLOPs Utilization (MFU) increased to 35\% during training, and the P99 tail latency was reduced to $\leq 400$ ms, enabling reliable large-scale inference.

Despite the relevance gains verified by human evaluation, an initial online deployment revealed a slight degradation in critical business metrics (e.g., orders and GMV). Root-cause analysis traced this issue to the upstream recall stage, which disproportionately retrieved items with strong semantic matching but low sales velocity, thereby reducing conversion efficiency. This observation highlights a common trade-off in e-commerce search: optimizing recall purely for semantic relevance can suppress commercially competitive items whose textual matches are less explicit.

To address this misalignment, we introduced a multi-channel recall mechanism that incorporates personalized signals, along with a pre-ranking stage that jointly optimizes relevance and commercial potential. We then conducted a 4-day online A/B test on 4\% traffic over eligible queries. The experiment achieved Direct Clean GMV +0.34\%, Direct Clean Orders -0.72\%, Direct Clean IPV +3.89\%, and Exposed PV +3.86\%. Overall, these results show consistent gains in user experience as measured by IPV and PV, with no material degradation of core business metrics. This underscores the necessity of co-designing recall and ranking to balance semantic quality and commercial objectives in end-to-end search systems.

\section{Conclusion}
We propose \textbf{TaoSR-AGRL}, an adaptive guided reinforcement learning framework that systematically addresses reward sparsity and exploration stagnation in LLM-based e-commerce search relevance. It integrates \emph{Rule-aware Reward Shaping}, which transforms sparse terminal outcomes into dense rewards to encourage faithful reasoning, and \emph{Adaptive Guided Replay}, which provides targeted, on-demand guidance to steer exploration toward high-value trajectories for challenging instances. Extensive offline and online evaluations show that TaoSR-AGRL consistently outperforms strong baselines, improving relevance accuracy and rule adherence while also boosting user engagement in real-world business settings. Deployed on Taobao serving hundreds of millions users, it provides a verified framework for applying reinforcement learning to build robust, industrial-scale reasoning systems.

\bibliographystyle{ACM-Reference-Format}
\balance
\bibliography{TaoSR-AGRL}

\clearpage
\appendix
\section{Ablation Study on Rule-aware Reward Shaping}
\label{sec:ablation_reward_shaping}

We conducted an ablation study to validate the design of our Rule-aware Reward Shaping. The study evaluates two key components: (1) the \textbf{validity gate} ($\mathbb{I}_{\mathrm{Gate}}$), which selectively applies rewards, and (2) the \textbf{reasoning reward} ($R_{\mathrm{reason}}$), which promotes logical consistency. TaoSR-AGRL employs a hard gate (Gating Coefficient $\lambda=0$), meaning no reward is assigned if the trajectory fails to satisfy the gating condition. To assess whether the model derives relevance in accordance with the operational rules, we report the Rule Adherence Rate (RAR). RAR measures the proportion of predictions where the final Relevance tier is consistent with the predicted Category and Attribute tiers under the Relevance Derivation Rules in Table~\ref{tab:appendix_derivation_rules}. Consequently, RAR quantifies rule-following behavior and complements standard performance metrics (F1, Accuracy): a model may achieve high Accuracy while violating internal rules, a discrepancy RAR is designed to reveal.

We compare against these variants:
\begin{itemize}
    \item \textbf{Soft Gating}: When the gate is closed (i.e., final answer is wrong), the reward is attenuated by a factor $\lambda\in\{0.2, 0.5, 0.8\}$ instead of being nullified.
    \item \textbf{w/o Gating}: The validity gate is removed and fine-grained rewards are always applied (equivalent to setting $\lambda=1$).
    \item \textbf{w/o $R_{\mathrm{reason}}$}: The reasoning reward component is removed; the final reward is the average of $R_{\mathrm{cate}}$ and $R_{\mathrm{attr}}$.
\end{itemize}

As shown in Table~\ref{tab:ablation_reward_shaping}, TaoSR-AGRL outperforms all ablated variants and attains the highest RAR. We observe three consistent trends: (1) relaxing or removing the gate causes \emph{reward leakage} into invalid trajectories, which not only degrades task metrics but also reduces RAR, indicating inferior rule adherence; (2) increasing the softness of the gate exacerbates this leakage and further diminishes internal consistency, underscoring the necessity of a hard gate to concentrate learning on valid reasoning chains; (3) removing $R_{\mathrm{reason}}$ yields notably lower RAR despite similar primary metrics, revealing shortcut solutions—a form of reward hacking—that bypass the intended logic. Collectively, these findings demonstrate that the hard gate provides a strict supervision boundary, while $R_{\mathrm{reason}}$ regularizes the solution space to ensure faithful, rule-consistent derivations, thereby improving both utility and reliability.

\begin{table}[h]
    \centering
    \caption{Ablation study on the components of our Rule-aware Reward Shaping on the In-the-Wild evaluation set. }
    \label{tab:ablation_reward_shaping}
    \begin{tabular}{lccc}
        \toprule
        \textbf{Method} & \textbf{F1 Score}  & \textbf{Accuracy}  & \textbf{RAR}\\
        \midrule
        Soft Gating ($\lambda=0.2$) & 60.54 & 73.96 & 84.07\\
        Soft Gating ($\lambda=0.5$) & 59.93 & 74.06 & 81.96\\
        Soft Gating ($\lambda=0.8$) & 59.77 & 73.83 & 83.58\\
        w/o Gating ($\lambda=1$) & 59.80 & 73.68 & 83.23\\
        w/o $R_{reason}$ & 61.28 & 74.02 & 79.84\\
        \midrule
        TaoSR-AGRL & \textbf{61.53} & \textbf{74.28} & \textbf{87.25}\\
        \bottomrule
    \end{tabular}
\end{table}

\section{Impact of the Replay Trigger Threshold}
\label{sec:appendix_tau}

Table~\ref{tab:tau_ablation} presents the sensitivity analysis of the replay trigger threshold $\tau$ on the In-the-Wild evaluation set. The table reports Macro F1 and Accuracy scores as $\tau$ varies within the range of $[0, 0.5]$, demonstrating that optimal performance is achieved at $\tau=0.1$.

\begin{table}[htbp]
    \centering
    \caption{Sensitivity analysis of the Replay Trigger Threshold $\tau$ on the In-the-Wild evaluation set.}
    \label{tab:tau_ablation}
    \begin{tabular}{ccc}
        \toprule
        \textbf{Threshold ($\tau$)} & \textbf{Macro F1} & \textbf{Accuracy} \\
        \midrule
        0 & 60.23 & 73.58 \\
        \textbf{0.1} & \textbf{61.53} & \textbf{74.28} \\
        0.2 & 59.91 & 74.10 \\
        0.3 & 59.40 & 73.77 \\
        0.4 & 59.45 & 73.64 \\
        0.5 & 58.70 & 74.27 \\
        \bottomrule
    \end{tabular}
\end{table}

\section{Relevance Derivation Rules}
\label{sec:ablation_rules}
Table~\ref{tab:appendix_derivation_rules} summarizes the \emph{Relevance Derivation Rules} currently employed on Taobao. Given specific Category and Attribute tiers, the corresponding relevance tier is deterministically mapped as shown.

\begin{table}[htbp]
\centering 
\caption{Relevance Derivation Rules.}
\label{tab:appendix_derivation_rules}
\begin{tabular}{@{}lccccc@{}}
\toprule
\diagbox{Cate}{Attr} & \textbf{Excellent} & \textbf{Related} & \textbf{Mismatch} & \textbf{Irrelevant}\\
\midrule
\textbf{Excellent}  & Excellent & Related   & Mismatch   & Irrelevant \\
\textbf{Related}    & Related   & Related   & Mismatch   & Irrelevant \\
\textbf{Mismatch}   & Mismatch  & Mismatch  & Mismatch   & Irrelevant \\
\textbf{Irrelevant} & Irrelevant& Irrelevant& Irrelevant & Irrelevant \\
\bottomrule
\end{tabular}
\end{table}

\section{Case Study}
We present two examples involving intricate e-commerce business logic to illustrate the effectiveness of Adaptive Guided Replay. By providing guidance tailored to the model's needs, the model successfully learns to explore and apply distinct reasoning rules, as demonstrated in Table~\ref{tab:case1} and Table~\ref{tab:case2}.

\onecolumn
\begin{table}[htbp]
\caption{Case Study 1: Impact of Adaptive Guided Replay}
\begin{center}
    \begin{tcolorbox}[colback=gray!5,colframe=black!50,title=Case Study: Impact of Adaptive Guided Replay on a query for \enquote{cashmere coat}.]
        \begin{tabular}{p{0.45\columnwidth} p{0.45\columnwidth}}
        \textbf{Original Prompt} & \textbf{Response} \\
        \hline
        Query: what is cashmere clothing? \newline
        Item: Women's Undyed Cashmere Coat ...
        &
        \textcolor{red}{4-Excellent}\newline
        1. Query: The user is looking for ...\newline
        2. Item: This cashmere coat ... \newline
        3. Category Match: ... The conclusion is Excellent.\newline
        4. Attribute Match: ... \textcolor{red}{Undyed Cashmere Coat contains cashmere} ... \textcolor{red}{The conclusion is Excellent}.\newline
        5. Judgment: ... Relevance label is 4-Excellent.
        \end{tabular}

        \bigskip

        \begin{tabular}{p{0.45\columnwidth} p{0.45\columnwidth}}
        \textbf{Prompt with Guidance} & \textbf{Response} \\
        \hline
        Query: what is cashmere clothing? \newline
        Item: Women's Undyed Cashmere Coat ... \newline
        \textbf{Guidance: Perform relevance reasoning per the guidance below, with Attribute and Relevance as Mismatch.}
        &
        \textcolor{green!60!black}{2-Mismatch}\newline
        1. Query: The user is looking for ...\newline
        2. Item: This cashmere coat ... \newline
        3. Category Match: ... The conclusion is Excellent.\newline
        4. Attribute Match: ... \textcolor{green!60!black}{Undyed Cashmere Coat contains cashmere, but the content is below 50\%} ... \textcolor{green!60!black}{The conclusion is Mismatch}.\newline
        5. Judgment: ... Relevance label is 2-Mismatch.
        \end{tabular}
    \end{tcolorbox}
\end{center}
  \label{tab:case1} 
\end{table}

\begin{table}[htbp]
 \caption{Case Study 2: Impact of Adaptive Guided Replay}
\begin{center}
    \begin{tcolorbox}[colback=gray!5,colframe=black!50,title=Case Study: Impact of Adaptive Guided Replay on a query for \enquote{chiffon dress}.]
        \begin{tabular}{p{0.45\columnwidth} p{0.45\columnwidth}}
        \textbf{Original Prompt} & \textbf{Response} \\
        \hline
        Query: recommend a chiffon dress suitable for a seaside vacation. \newline
        Item: French-style Floral Vacation Dress ...
        &
        \textcolor{red}{2-Mismatch}\newline
        1. Query: The user is looking for ...\newline
        2. Item: This floral vacation dress ... \newline
        3. Category Match: ... The conclusion is Excellent.\newline
        4. Attribute Match: ... \textcolor{red}{The dress contains chiffon, but the content is less than 50\%} ... \textcolor{red}{The conclusion is Mismatch}.\newline
        5. Judgment: ... Relevance label is 2-Mismatch.
        \end{tabular}

        \bigskip

        \begin{tabular}{p{0.45\columnwidth} p{0.45\columnwidth}}
        \textbf{Prompt with Guidance} & \textbf{Response} \\
        \hline
        Query: recommend a chiffon dress suitable for a seaside vacation. \newline
        Item: French-style Floral Vacation Dress ... \newline
        \textbf{Guidance: Perform relevance reasoning per the guidance below, with Attribute and Relevance as Excellent.}
        &
        \textcolor{green!60!black}{4-Excellent}\newline
        1. Query: The user is looking for ...\newline
        2. Item: This floral vacation dress ... \newline
        3. Category Match: ... The conclusion is Excellent.\newline
        4. Attribute Match: ... \textcolor{green!60!black}{The dress contains chiffon} ... \textcolor{green!60!black}{The conclusion is Excellent}.\newline
        5. Judgment: ... Relevance label is 4-Excellent.
        \end{tabular}
    \end{tcolorbox}
\end{center}
 \label{tab:case2} 
\end{table}

\end{document}